\documentclass[a4paper,12pt]{article}
\usepackage[top=3cm,bottom=3.4cm,left=2cm,right=2cm]{geometry}

\usepackage{bm}
\usepackage{amsmath}
\usepackage{amssymb}
\usepackage{cite}

\newcommand{\Tc}{{\tilde c}}
\newcommand{\Bpartial}{{\overline{\partial}}}
\newcommand{\Bz}{{\bar z}}

\renewcommand{\thepage}{}
\makeatletter
\@addtoreset{equation}{section}
\renewcommand{\theequation}{\thesection.\@arabic\c@equation}
\makeatother
\renewcommand{\thefootnote}{\fnsymbol{footnote}}

\begin{document}
\begin{titlepage}
\title{
\vspace*{-4ex}
\hfill
\begin{minipage}{3.5cm}
\end{minipage}\\
 \bf 
Two-point String Amplitudes Revisited \\
by Operator Formalism
\vspace{0.5em}
}

\author{
Shigenori {\sc Seki}\footnote{\tt sseki@mail.doshisha.ac.jp}\medskip\\ 
{\it Faculty of Sciences and Engineering, Doshisha University,}\\ 
{\it 1-3 Tatara-Miyakodani, Kyotanabe, Kyoto 610-0394, Japan}\medskip \\
{\it Osaka City University Advanced Mathematical Institute (OCAMI),}\\
{\it 3-3-138, Sugimoto, Sumiyoshi-ku, Osaka 558-8585, Japan}\bigskip \\
and\bigskip\\
Tomohiko {\sc Takahashi}\footnote{\tt tomo@asuka.phys.nara-wu.ac.jp}\medskip\\
{\it Department of Physics, Nara Women's University,}\\
{\it Nara 630-8506, Japan}
}
\date{}
\maketitle

\begin{abstract}
\normalsize
So far we have considered that a two-point string amplitude vanishes due to the infinite volume of 
residual gauge symmetry. 
However recently Erbin-Maldacena-Skliros have suggested that 
the two-point amplitude can have non-zero value, because one can cancel the infinite volume 
by the infinity coming from on-shell energy conservation. 
They derived the two-point function by Fadeev-Popov method. 
In this paper we revisit this two-point string amplitude in the operator formalism. 
We find the mostly BRST exact operator which yields non-zero two-point amplitudes.
\end{abstract}

\vfill\noindent
\date{30 September 2019}
\end{titlepage}

\renewcommand{\thepage}{\arabic{page}}
\renewcommand{\thefootnote}{\arabic{footnote}}
\setcounter{page}{1}
\setcounter{footnote}{0}

\section{Introduction}

It has been shown that two-point tree level string amplitudes in flat space
are given by the free particle expression in conventional quantum field theory
\cite{Erbin:2019uiz}.
In open string theory, a two-point string amplitude in tree level can be
evaluated by a correlation function of two vertex operators inserted in 
the real axis of the upper-half plane. The automorphism of the 
upper-half plane, $\mathrm{PSL}(2,\mathbb{R})$, is partially fixed by
the two points, but the gauge volume of the residual symmetry becomes
infinity. Usually, the resulting two-point amplitude was argued to
vanish due to the infinite volume.
However, in Ref.~\cite{Erbin:2019uiz}, it is pointed out that 
this infinity cancels $\delta(0)$ arising from
the energy conservation of the amplitude in the path integral formulation, 
so that the standard free particle expression is given
for the two string amplitudes. 

The tree two-point amplitude has
been usually thought to vanish also in 
the covariant operator formalism
based on the BRST symmetry, while it provides a fascinating approach to
evaluating string amplitudes.
In this formalism,
an $n$-point tree-level open string amplitude ($n\geq 3$) is constructed
by $n$ ghost operator insertions attached to the vertex operators and
$n-3$ anti-ghost operator insertions associated to the integration on
moduli space \cite{Polchinski}. 
Applying this rule to the two-point amplitude,
the ghost number gained by the vertex operators is two, which is lesser than 
the number of the ghost zero modes. 
Thus the two-point amplitude does not vanish only if another operator of ghost
number one is inserted.

In order to reproduce non-vanishing two-point amplitudes in the operator
formalism,
it is necessary to find a new type of vertex operators supplying an
additional ghost number in the amplitude.
Moreover, $\delta(0)$ given by the energy conservation should disappear
owing to this novel vertex operator.

The energy conservation factor $\delta(0)$ is expressed as
\begin{align}
\langle\, 0\, |\,{p_1}^0-{p_2}^0 \,\rangle 
	=2\pi \delta({p_1}^0-{p_2}^0)\,, \quad
{p_i}^0=\sqrt{{\bm{p}_i}^2+m^2} \quad (i=1,\,2)
\end{align} 
where $\langle\,0\,|$ is the $\mathrm{SL}(2,\mathbb{R})$ invariant vacuum 
and $|\,p^0\,\rangle$ is an energy eigenstate.
This delta function provides $\delta(0)$ due to the momentum conservation,
 $\bm{p}_1=\bm{p}_2$.
Here we introduce the eigenstate $|\,x^0\,\rangle$
of the zero mode of the world-sheet field $X^0$ and
we replace $\langle\,0\,|$ by $\langle\,\tilde{0}\,| \equiv \langle\, x^0=0 \,|$. 
Then the energy conservation $\delta(0)$
changes to a finite value:
\begin{align}
\langle\,\tilde{0}\,|\,{p_1}^0-{p_2}^0\,\rangle
=e^{i0\cdot({p_1}^0-{p_2}^0)}=1 \,.
\end{align}
The operator corresponding to $|\,\tilde{0}\,\rangle$ is given by
\begin{align}
 \int_{-\infty}^\infty \frac{dq}{2\pi}\,e^{iq X^0(z,\bar{z})} \,.
\end{align}
Hence, there is a possibility that the two-point string amplitudes is provided in
the operator formalism, if this operator can be rewritten as a BRST
invariant operator with the ghost number one.

The purpose of this paper is to derive such an operator and then to
show that the two-point string amplitudes can be calculated also within the
operator formalism. We also discuss the the Lorentz and conformal
invariance and some applications.

\section{Two-point amplitudes in the operator formalism}

\subsection{Open strings} \label{sec:openstrings}

Let us consider open strings in $D$ flat spacetime directions.
The world-sheet is given by
the upper half plane and the open string boundary corresponds to the real
axis.

We introduce the following operator made of the 0-th string coordinate
$X^0(z,\bar{z})$ and the ghost field $c(z)$:
\begin{align}
 {\cal V}_0(z) \equiv \int_{-\infty}^\infty
\frac{dq}{\pi \alpha' i}\Bigl(c\partial X^0 e^{-iqX^0}
-i\alpha' q (\partial c)e^{-iqX^0}\Bigr) \,,
\label{eq:V0}
\end{align}
where the position $z$ is along the real axis.

One can easily check that ${\cal V}_0(z)$ is a BRST invariant operator. 
Moreover, the operator can be rewritten as
\begin{align}
 {\cal V}_0(z)= \int_{-\infty}^\infty
\frac{dq}{\pi \alpha' q}\,\bigl[\,Q_{\rm B},\, e^{-iqX^0}\,\bigr] \,,
\label{eq:V0Qb}
\end{align} 
where $Q_{\rm B}$ is the BRST operator.
This implies that ${\cal V}_0(z)$ is a {\it mostly} BRST exact operator. 
However, the integrand is ill-defined at $q=0$ since it is given as $0/0$. 
Therefore, the integrand is expected to behave like a
distribution supported at $q=0$.\footnote{From Eq.~\eqref{eq:V0Qb}, the operator
${\cal V}_0(z)$ is formally expressed by using
the step function of $X^0(z)$, {\it i.e.}, 
${\cal V}_0(z)= -2i \alpha'^{-1}\bigl[\, Q_{\rm B},\,\theta\bigl(X^0(z)\bigr)\,\bigr]$.
It is reminiscent of the picture changing operator
in superstring theory \cite{IK}, which is given by the commutator of $Q_{\rm B}$
with the step function of the $\beta$
ghost \cite{Polchinski,Friedan:1985ge}.}

Here we consider the correlation function including ${\cal V}_0$:
\begin{align}
{\cal A}= \langle\, 0 \, |\, {\cal V}_0(z_0){\cal V}_1(z_1){\cal V}_2(z_2) \,|\, 0 \,\rangle\,,
\label{eq:correlfunc}
\end{align} 
where all $z_i$ are on the real axis and 
the two vertex operators are given as ${\cal V}_i(z)=c V_i(z)$ ($i=1,2$)
for the matter vertex $V_i(z)$ with conformal weight one, 
the momentum ${p_i}^\mu$ ($\mu=0,\cdots,D-1$) and the mass $m_i$.  
By the energy-momentum conservation, the correlation function is expressed
as
\begin{align}
 {\cal A}=\int_{-\infty}^\infty dq\, (2\pi)^{D-1}\delta(q-{p_1}^0+{p_2}^0)
\delta^{D-1}(\bm{p}_1-\bm{p}_2)\times \cdots \,,
\end{align}
where the on-shell condition is satisfied: ${p_i}^0=\sqrt{{\bm{p}_i}^2+{m_i}^2}$.
If ${m_1}^2\neq {m_2}^2$, we find ${p_1}^0-{p_2}^0\neq 0$ and so
$q$ is replaced by a non-zero value after the $q$-integration.
In this case, since the integrand of ${\cal V}_0$ is a BRST exact operator
for $q\neq 0$, the correlation function turns out to be zero.

In the case of ${m_1}^2={m_2}^2$, the delta function for the energy conservation 
becomes $\delta(q)$. 
Therefore, taking out of the contribution of $q=0$ after the $q$-integration, 
the correlation function \eqref{eq:correlfunc} can be rewritten as
\begin{align}
 {\cal A}= \langle\, 0 \,|\,c\partial X^0(z_0)\,cV_1(z_1)\,cV_2(z_2)\,|\,0\,\rangle'\,,
\label{eq:opencorrel}
\end{align} 
where the prime means that it does not include the delta function
of the energy conservation, which vanishes due to the $q$-integration.
This expression is equal to the two-point string amplitude derived by
using the Fadeev Popov method in Ref.~\cite{Erbin:2019uiz}.
Consequently, we can find that the correlation function
provides the standard free particle expression of the two-point amplitude:
\begin{align}
 {\cal A}=2 p^0 (2\pi)^{D-1}\delta^{D-1}(\bm{p}_1-\bm{p}_2) \,,
\label{eq:twoptamp}
\end{align}
where $p^0 = {p_1}^0 = {p_2}^0$. If the mass eigenstates are degenerate,
there is an extra Kronecker delta for the orthonormal basis.

Here we should give some comments.
The first term appeared in the right hand side of Eq.~\eqref{eq:V0}
plays an essential role to give a non-zero result of
the correlation function. Actually, the operator ${\cal V}_0(z)$
is written as
\begin{align}
 {\cal V}_0(z)
= -{2i \over \alpha'} \bigl\{c\partial X^0(z)\,\delta\bigl(X^0(z)\bigr)
	+ \alpha' \partial c(z)\,\delta'\bigl(X^0(z)\bigr) \bigr\} \,,
\end{align}
and the first term is exactly equal to the expression given in Ref.~\cite{Erbin:2019uiz}; 
the delta function related to the gauge fixing condition $X^0=0$ 
and the associated Fadeev-Popov determinant.
The second term is also indispensable for realization of
correct two-point amplitudes. If the second term is absent,
the BRST exactness for $q\neq 0$ is not be assured and so
two-point amplitudes would have non-zero values even in the different
mass case.

The right-hand side of Eq.~\eqref{eq:twoptamp} is the standard inner product
for on-shell states in quantum field theories. In Ref.~\cite{Polchinski},
it is pointed out that this standard one is not correctly given by the
conventional inner products in which the ghost zero mode $c_0$ is
inserted. Instead, a reduced inner product is defined by ignoring $X^0$
and ghost zero modes to provide the inner product for on-shell
states \cite{Polchinski}.
We emphasize that the standard inner product for on-shell states is
given by a correlation function (\ref{eq:correlfunc}), in which the
inner product is conventional including $c_0$ insertion.

By state-operator correspondence, the vertex operator ${\cal V}_0(z)$
corresponds to the state $|\,{\cal V}_0\,\rangle = {\cal V}_0(0)|\,0\,\rangle$.
This state satisfies the subsidiary condition 
$Q_{\rm B}\, |\,{\cal V}_0\,\rangle=0$ \cite{Kato:1982im,Kugo:1979gm},
but does not satisfy the additional condition 
$b_0\,|\,{\cal V}_0\,\rangle=0$ \cite{Kato:1982im},
since the second term in the right hand side of Eq.~\eqref{eq:V0} corresponds to a state
proportional to $c_0$. Therefore, ${\cal V}_0(z)$ is not an on-shell state
and so outside the conventional Hilbert space of physical states.

\subsection{Lorentz and conformal invariances}

The operator (\ref{eq:V0}) is a novel vertex operator to provide the
two-point amplitude in the operator formalism. However, this expression
(\ref{eq:V0}) breaks the Lorentz invariance since it picks up the time
direction as special. First, we will discuss why
the Lorentz invariant amplitude (\ref{eq:twoptamp}) is obtained by using
the Lorentz non-invariant operator (\ref{eq:V0}).

Let us consider the infinitesimal Lorentz transformation:
\begin{align}
 X^\mu \,\rightarrow\, {X'}^\mu=X^\mu +{\epsilon^\mu}_\nu\,X^\nu \,,
\end{align}
where $\epsilon^{\mu\nu}$ is an anti-symmetric infinitesimal parameter.
For the infinitesimal transformation, the variation of the operator
(\ref{eq:V0}) is expressed as
\begin{align}
 \delta{\cal V}_0=
\bigl[\,Q_{\rm B},\,\int_{-\infty}^\infty \frac{dq}{\pi \alpha' i}\,
{\epsilon^0}_\nu X^\nu e^{-iqX^0}\,\bigr] \,.
\label{eq:LTV0}
\end{align}
The important point here is that the factor $1/q$ does not appear in
contrast to the expression (\ref{eq:V0Qb}). Since the integrand
in the above has no singularity for any $q$, the infinitesimal variation
is given by a BRST exact operator.
Therefore, by finite Lorentz transformations, 
the operator ${\cal V}_0$ is transformed as
\begin{align}
 {\cal V}_0 \,\rightarrow\, {\cal V}_0 +[\,Q_{\rm B},\,\bullet\,] \,.
\end{align}
The BRST exact term does not contribute to the correlation function for
physical operators. Consequently, the amplitude satisfies Lorentz invariance,
although the operator (\ref{eq:V0}) is apparently breaks Lorentz
invariance.

In the same manner, we find conformal invariance of the amplitude
although ${\cal V}_0$ is not a conformal invariant operator.
For an infinitesimal conformal transformation $\delta z=\epsilon(z)$,
the operator ${\cal V}_0(z)$ is transformed to
\begin{align}
 \delta_\epsilon {\cal V}_0 = 
 \bigl[\,Q_{\rm B},\, \int_{-\infty}^\infty \frac{dq}{\pi \alpha' i} \Bigl(
\epsilon \partial X^0 e^{-iqX^0} -i\alpha' q (\partial \epsilon )e^{-iqX^0}
\Bigr) \,\bigr] \,.
\end{align}
Similar to Eq.~\eqref{eq:LTV0}, the integrand includes no poles of $q$ 
and so the infinitesimal variation is given as a well-defined BRST exact operator.
Therefore, the amplitude is provided as a conformal invariant quantity
as well as the case of the Lorentz invariance.

\subsection{Closed strings}

Let us straightforwardly apply the open string ${\cal V}_0$ to 
a closed string one, 
\begin{align}
{\cal V}_0(z,\Bz) &\equiv 2\int_{-\infty}^\infty {dq \over \pi \alpha' q}\,
	\bigl[\,{\cal Q}_{\rm B},\, e^{-iqX^0} \,\bigr] \nonumber\\
&= 2\int_{-\infty}^\infty {dq \over \pi \alpha' i} \biggl( \bigl(c\partial X^0 +\Tc \Bpartial X^0\bigr)e^{-iqX^0}
	-{i\alpha' q \over 4} \bigl(\partial c +\Bpartial \Tc\bigr) e^{-iqX^0} \biggr) \,,
\end{align}
where ${\cal Q}_{\rm B}$ is the BRST operator, ${\cal Q}_{\rm B} = Q_{\rm
B} + {\tilde Q}_{\rm B}$. 
The factor 2 is due to the left and right movers of closed string. 
This is also a mostly BRST exact operator 
like the open string ${\cal V}_0$ defined by Eq.~\eqref{eq:V0}.
However, it is not straightforward to apply the correlation function 
of open strings \eqref{eq:opencorrel} to the one of closed strings. 
For the two vertex operators $V_i$ ($i=1,2$) with conformal weight $(1,1)$, 
one may easily guess the correlation function on a sphere as 
$\langle\, 0 \,|\, {\cal V}_0(z_0,\Bz_0)\, c\Tc V_1(z_1,\Bz_1)\, c\Tc V_2(z_2,\Bz_2) \,|\,0\rangle$ 
in a standard manner, where $|\,0\,\rangle$ is 
the $\mathrm{SL}(2,\mathbb{C})$ invariant vacuum.
However it vanishes, because its total ghost number is five, which
mismatches to six, the number of zero modes on the sphere.
In order for the ghost number matching, 
we try the following assignment of ghosts:
\begin{align}
{\cal V}_1(z,\Bz) = (-c\partial c)\Tc V_1\,, \quad 
{\cal V}_2(z,\Bz) = c\Tc V_2 \,. 
\end{align}
Note that ${\cal V}_1$ is also BRST invariant.
Then the closed string amplitude,
\begin{align}
{\cal A} = \big< 0\big| {\cal V}_0(z_0,\Bz_0){\cal V}_1(z_1, \Bz_1) {\cal V}_2(z_2,\Bz_2) \big|0\big> \,,
\label{eq:twoptampclosed}
\end{align}
becomes the two-point amplitude of the standard free particles. 
Indeed we calculate the amplitude \eqref{eq:twoptampclosed} in the equal-mass case, 
\begin{align}
{\cal A} &= \langle\, 0 \,|\, \Tc \Bpartial X^0 (\Bz_0)\, (-c \partial c) \Tc V_1(z_1,\Bz_1)\, c\Tc V_2(z_2,\Bz_2) \,|\,0\,\rangle' \nonumber\\
&= 2 p^0 (2\pi)^{D-1}\delta^{D-1}(\bm{p}_1-\bm{p}_2) \,. \label{eq:resultclosed}
\end{align}
The matter correlation $\langle \Bpartial X^0 V_1 V_2\rangle$ includes 
a factor $z_{12}^{-2} \Bz_{01}^{-1} \Bz_{02}^{-1} \Bz_{12}^{-1}$. 
Its anti-holomorphic part is canceled by the ghost correlation 
$\langle \Tc(z_0) \Tc(z_1) \Tc(z_2) \rangle$, 
while its holomorphic part is canceled by $\langle -c\partial c(z_1) c(z_2)\rangle$. 
This is one of the reasons why we attach the ghost fields $c\partial c$ to the vertex operator $V_1$. 
It is also possible to assign the anti-ghost fields $\Tc \Bpartial \Tc$ to the vertex operator as 
${\cal V}_1 = c(-\Tc\Bpartial\Tc)V_1$ and ${\cal V}_2 = c\Tc V_2$. 
However these ghost assignments cause a problem, that is, 
$(-c\partial c) {\tilde c}$ or $c (-{\tilde c} \Bpartial{\tilde c})$ 
are not annihilated by $b_0 -{\tilde b}_0$. 
Such annihilation is necessary for a physical closed string vertex operator \cite{Polchinski:1988jq,rf:PN,Sen:2014pia}. 
Even if we take a liner combination of these assignments, 
the annihilated condition is satisfied but the resulting amplitude becomes vanishing.
All in all, currently the naive application of the open string to the closed string is not successful.

\section{Concluding remarks}

We have studied the two-point string amplitude in the covariant operator formalism. 
For open strings we have found the mostly BRST exact operator ${\cal V}_0$ which yields the non-zero 
two-point amplitude. 
Although the operator ${\cal V}_0$ itself violates Lorentz and conformal invariances, 
the two-point amplitude recovers them. 

For open strings we have obtained the two-point amplitude like a standard free particles \eqref{eq:twoptamp}, 
which is the same amplitudes that Ref.~\cite{Erbin:2019uiz} derived by the use of Fadeev-Popov method. 
In our formalism the amplitude is given by the correlation function of ${\cal V}_0$ and two vertex operators $V_i$ with the ghost field $c$. 
On the other hand, 
for closed strings, we have introduced the mostly BRST exact operator ${\cal V}_0$ 
in the same way as open strings, but the correlation function of ${\cal V}_0$ and the two vertex operators 
$V_i$ with the ghost fields $c\Tc$ does not make sense 
for lack of the total ghost number of this correlation function. 
Therefore, we have tried to assign the ghost fields $(-c\partial c)\Tc$ 
on one of the vertex operators $V_i$ and $c \Tc$ on the other. 
This provides us with the non-zero two-point closed string amplitude \eqref{eq:resultclosed}, 
but there appears the problem, {\it i.e.}, $(-c\partial c)\Tc$ or $c (-\Tc\Bpartial \Tc)$ 
are not annihilated by $b_0 - {\tilde b}_0$. 
Since the simple application of the open string amplitude for the closed string one does not work, 
we have to find a prescription for the closed string case. 

The mostly BRST exact operator ${\cal V}_0$ plays a significant role in our formulation. 
In particular, the BRST quantization of strings is valuable for formulating string field theories.

Following our formulation, it can be straightforward to consider the two-point amplitude of 
open superstrings. On the other hand, for closed superstrings, we may be confronted 
with a difficulty similar to the closed bosonic strings \cite{ST}. 

It should be commented that the insertion of ${\cal V}_0$ is expected to
implement a special choice of gauge fixing of the dilatation subgroup
$\mathrm{SL}(2,\mathbb{R})$. Actually, ${\cal V}_0$ is a BRST invariant extension
of the operator given by the Fadeev-Popov method in
Ref.~\cite{Erbin:2019uiz}, as pointed
out in the section \ref{sec:openstrings}. So, it is expected to be possible to compute
arbitrary tree-level $n$-point amplitudes by inserting ${\cal V}_0$.
For example, a three-point amplitude may be obtained by a correlator of
${\cal V}_0$, ${\cal V}_i$ $(i=1,2)$ at specific points, which fix
$\mathrm{SL}(2,\mathbb{R})$, and $\int dz_3\, V_3(z_3)$, where $V_3$ is a matter vertex
operator with dimension one.
Although the $q$-integration in ${\cal V}_0$ eliminates the
energy-conserving delta function, the moduli integration of $z_3$ may
supply another energy-conserving delta function to lead the correct
three-point amplitude. We will provide the
details of this study in the near future \cite{ST}.

\section*{Acknowledgments}
The authors would like to thank Isao Kishimoto for valuable comments.
We also thank the Galileo Galilei Institute for Theoretical
Physics and INFN for hospitality and support during the workshop ``String
Theory from a Worldsheet Perspective''.
S.~S. was supported in part by JSPS Grant-in-Aid for Scientific Research
(C) \#17K05421 and by MEXT Joint Usage/Research Center on Mathematics
and Theoretical Physics at OCAMI.
The research of T.~T. was supported in part by Nara Women's University
Intramural Grant for Project Research.


\end{document}